\documentclass[
 aip,       
 amsmath,amssymb,
 twocolumn,  
]{revtex4}

\usepackage{textcase}
\usepackage{graphicx}
\usepackage{dcolumn}
\usepackage{bm}
\usepackage{amsmath,amsthm,amssymb,mathrsfs}

\begin{document}

\title{Dependence of spin torque diode voltage on applied field direction}

\author{Tomohiro Taniguchi}
\author{Hiroshi Imamura}

\affiliation{ 
Spintronics Research Center, AIST, 1-1-1 Umezono, Tsukuba 305-8568, Japan\\
}

\date{\today}%

\begin{abstract}
      The optimum condition of an applied field direction 
      to maximize spin torque diode voltage 
      was theoretically derived 
      for a magnetic tunnel junction with a perpendicularly magnetized free layer 
      and an in-plane magnetized pinned layer. 
      We found that the diode voltage for a relatively small applied field is maximized 
      when the projection of the applied field to the film-plane is 
      parallel or anti-parallel to the magnetization of the pinned layer. 
      However, by increasing the applied field magnitude, 
      the optimum applied field direction shifts from the parallel or anti-parallel direction. 
      These analytical predictions were confirmed by numerical simulations. 
\end{abstract}

\maketitle


\section{Introduction}
\label{sec:Introduction}

Magnetization dynamics induced by spin torque 
in nano-structured ferromagnets \cite{slonczewski89,slonczewski96,slonczewski02,berger96} 
have provided interesting phenomena, 
such as magnetization switching and oscillation. 
Many spintronics devices utilizing the spin torque have been proposed, 
such as, a magnetic random access memory (MRAM) based on magnetic tunnel junctions (MTJs) 
and a microwave oscillator \cite{yuasa08,suzuki08}. 
The spin torque diode effect \cite{tulapurkar05,kubota08,sankey08,suzuki09,petit07,fuchs07,chen09,yakata09,wang09,boone09,ishibashi10,cheng10,miwa12,bang12,zhu12} 
is also an important phenomenon, 
in which an alternating current applied to an MTJ is rectified 
by synchronizing the resonant oscillation of tunnel magnetoresistance (TMR) \cite{yuasa04,parkin04} 
by spin torque with the alternating current. 
The spin torque diode effect has been used 
to quantitatively evaluate the strength of the spin torque \cite{kubota08,sankey08}. 


The spin torque diode effect is applicable to a magnetic sensor application, 
where a small magnetic field from a ferromagnetic or paramagnetic particle 
modulates the resonance condition of the spin torque diode \cite{miwa12}. 
In such a sensor application, 
a large diode voltage is required to enhance sensitivity, 
defined as the ratio 
between the input power and the diode voltage \cite{ishibashi10}. 
It should also be noted that 
the direction of the applied field, 
which is proportional to the spin of the particle, 
points in an arbitrary direction. 
Thus, it is important to clarify the relation between 
the spin torque diode voltage and the applied field direction, 
and to maximize the spin torque diode voltage. 


In this paper, 
we derive the optimum condition of the applied field direction 
to maximize the spin torque diode voltage of 
an MTJ with a perpendicularly magnetized free layer 
and an in-plane magnetized pinned layer. 
This type of MTJ was recently developed in experiments \cite{yakata09,kubota12,kubota13}, 
and is considered an ideal candidate for spin torque diode application 
because of its narrow linewidth and high diode voltage. 
We first derived the general formula of the spin torque diode voltage, 
and then, applied the formula to the system underconsideration. 
The main result is Eq. (\ref{eq:p_z_opt_perp}), 
which represents the applied field direction at the maximized diode voltage. 
The diode voltage for a relatively small applied field is maximized 
when the projection of the applied field to the film-plane is 
parallel or anti-parallel to the magnetization of the pinned layer. 
However, 
the optimum applied field direction shifts from the parallel or anti-parallel direction 
by increasing the applied field magnitude. 
These results are confirmed numerically. 


The paper is organized as follows. 
In Sec. \ref{sec:Solution to the linearized LLG equation}, 
we derive the analytical solution to the linearized Landau-Lifshitz-Gilbert (LLG) equation 
of the free layer. 
In Sec. \ref{sec:Spin torque diode voltage}, 
the general formula of the spin torque diode voltage 
and its dependence on the magnetization alignment are discussed. 
Section \ref{sec:Optimum condition of applied field direction} is 
the main section in this paper, 
where we derive the optimum condition for the applied field direction 
in an MTJ with a perpendicularly magnetized free layer 
and an in-plane magnetized pinned layer. 
Section \ref{sec:Conclusions} is devoted to the conclusions. 


\section{Solution to the linearized LLG equation}
\label{sec:Solution to the linearized LLG equation}

\begin{figure}
\centerline{\includegraphics[width=0.5\columnwidth]{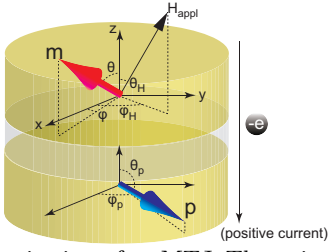}}\vspace{-3.0ex}
\caption{
         Schematic view of an MTJ. 
         The unit vectors pointing to the magnetization directions of the free and the pinned layers are denoted as 
         $\mathbf{m}$ and $\mathbf{p}$, respectively. 
         The positive current is defined as the electron flow from the pinned to the free layer. 
         \vspace{-3ex}}
\label{fig:fig1}
\end{figure}

In this section, we solve the linearized LLG equation 
for an arbitrary magnetization alignment. 
The system we consider is schematically shown in Fig. \ref{fig:fig1}, 
where the MTJ consists of free and pinned layers separated by a thin nonmagnetic spacer. 
The $x$, $y$, and $z$ axes are parallel to 
the uniaxial anisotropy axes of the free layer. 
The unit vectors pointing in the direction of the magnetizations 
of the free and the pinned layers are denoted as 
$\mathbf{m}$ and $\mathbf{p}=(\sin\theta_{\rm p}\cos\varphi_{\rm p},\sin\theta_{\rm p}\sin\varphi_{\rm p},\cos\theta_{\rm p})$, respectively, 
where the zenith and the azimuth angles of the magnetization of the pinned layer are denoted as $\theta_{\rm p}$ and $\varphi_{\rm p}$, respectively. 
We assume that the magnetization dynamics in the presence of the spin torque are well described by 
the macrospin LLG equation \cite{slonczewski89,slonczewski96,slonczewski02,berger96,landau35,lifshitz80,gilbert04}: 
\begin{equation}
\begin{split}
  \frac{d \mathbf{m}}{dt}
  =&
  -\gamma
  \mathbf{m}
  \times
  \mathbf{H}
  -
  \gamma
  a_{J}
  \mathbf{m}
  \times
  \left(
    \mathbf{p}
    \times
    \mathbf{m}
  \right)
\\
  &+
  \gamma
  b_{J}
  \mathbf{p}
  \times
  \mathbf{m}
  +
  \alpha
  \mathbf{m}
  \times
  \frac{d\mathbf{m}}{dt},
  \label{eq:LLG}
\end{split}
\end{equation}
where $\gamma$ and $\alpha$ are the gyromagnetic ratio and the Gilbert damping constant, respectively. 
The magnetic field $\mathbf{H}$ is defined as 
the derivative of the energy density $E$ with respect to the magnetization, 
i.e., 
\begin{equation}
  \mathbf{H}
  =
  -\frac{1}{M}
  \frac{\partial E}{\partial \mathbf{m}},
  \label{eq:field_H}
\end{equation}
where $M$ is the saturation magnetization. 
The energy density $E$ is given by 
\begin{equation}
\begin{split}
  E
  =&
  -M H_{\rm appl}
  \left[
    \sin\theta_{H}
    \sin\theta
    \cos(\varphi-\varphi_{H})
    +
    \cos\theta_{H}
    \cos\theta
  \right]
\\
  &+
  \sum_{\ell=x,y,z}
  2\pi M^{2}
  \tilde{N}_{\ell}
  m_{\ell}^{2},
  \label{eq:energy}
\end{split}
\end{equation}
where $H_{\rm appl}$, $\theta_{H}$, and $\varphi_{H}$ in the first term are 
the magnitude, the zenith angle, and the azimuth angle of the applied field, respectively. 
The second term of Eq. (\ref{eq:energy}) describes 
the uniaxial anisotropy energy. 
The coefficient $\tilde{N}_{\ell}$ ($\ell=x,y,z$) is defined as 
$4\pi M \tilde{N}_{\ell}=4\pi M N_{\ell}-H_{{\rm K}\ell}$, 
where $4\pi M N_{\ell}$ and $H_{{\rm K}\ell}$ are 
the shape anisotropy field (demagnetization field) and 
the crystalline anisotropy field along the $\ell$-axis, respectively. 
The demagnetization coefficients satisfy $N_{x}+N_{y}+N_{z}=1$. 


The two components of the spin torque in Eq. (\ref{eq:LLG}), 
the Slonczewski torque and the field like torque, are denoted as 
$a_{J}$ and $b_{J}$, respectively, 
whose explicit forms are given by 
\begin{equation}
  a_{J}
  =
  \frac{\hbar gI}{2eMV},
  \label{eq:a_J}
\end{equation}
and $b_{J}=\beta a_{J}$. 
Here, $I$ is the current and $V$ is the volume of the free layer, 
respectively. 
The positive current corresponds to the electron flow from the free to the pinned layer. 
We assume that both the direct (dc) and alternating (ac) currents are applied to the MTJ, 
i.e., $I=I_{\rm dc}+I_{\rm ac}(t)$. 
Thus, $a_{J}$ and $b_{J}$ are decomposed into the dc and the ac parts 
as $a_{J}=a_{J({\rm dc})}+a_{J({\rm ac})}$ and $b_{J}=b_{J({\rm dc})}+b_{J({\rm ac})}$, respectively. 
The magnitude of the direct current is on the order of 0.1-1.0 mA, 
while that of the alternating current is 0.1 mA \cite{tulapurkar05,kubota08}. 
As shown below, 
the present formula is valid for $|a_{J({\rm dc})}| < \alpha |\mathbf{H}|$, 
where the Gilbert damping constant is on the order of $10^{-2}$ \cite{oogane06}. 
The ratio of the Slonczewski torque to the field like torque, $\beta$, 
is on the order of 0.1 for MTJs \cite{tulapurkar05,kubota08,sankey08}. 
The factor $g$ characterizes the spin polarization of the current, 
and is given by \cite{slonczewski89,slonczewski96,slonczewski02} 
\begin{equation}
  g
  =
  \frac{\eta}{1+\lambda \mathbf{m}\cdot\mathbf{p}}.
  \label{eq:factor_g}
\end{equation}
The dimensionless parameters, $\eta$ and $\lambda$, 
characterize the magnitude of the spin polarization 
and the dependence of the spin torque 
on magnetization alignment, respectively. 
Although the relation among $\eta$, $\lambda$, and the material parameters depends 
on the theoretical models, 
the form of Eq. (\ref{eq:factor_g}) is applicable 
to spin torque in both MTJs and giant-magnetoresistive system \cite{xiao04}. 
For example, in the ballistic transport theory in MTJ, 
$\eta$ is proportional to the spin polarization of the density of state of the free layer 
and $\lambda = \eta^{2}$ \cite{slonczewski89,suzuki08,slonczewski05}. 
Below, we set $\lambda=0$ for simplicity. 
The spin torque diode voltage and its optimum condition with finite $\lambda$ are 
discussed in Appendix A. 


The solution to the LLG equation is derived 
in an $XYZ$-coordinate 
in which the $Z$-axis is parallel to the steady state of the magnetization of the free layer. 
We denote $\mathbf{m}$ at the steady state as $\mathbf{m}^{(0)}$, 
where the condition ${\rm d}\mathbf{m}^{(0)}/{\rm d}t=\bm{0}$ 
can be expressed in terms of 
the zenith and the azimuth angles, $(\theta,\varphi)$, as 
\cite{taniguchi13,vonsovskii66}: 
\begin{equation}
\begin{split}
  &
  H_{\rm appl}
  \left[
    \sin\theta_{H}
    \cos\theta
    \cos(\varphi-\varphi_{H})
    -
    \cos\theta_{H}
    \sin\theta
  \right]
\\
  &-
  4\pi M 
  \left(
    \tilde{N}_{x}
    \cos^{2}\varphi
    +
    \tilde{N}_{y}
    \sin^{2}\varphi
    -
    \tilde{N}_{z}
  \right)
  \sin\theta
  \cos\theta
\\
  &-
  a_{J({\rm dc})}
  \sin\theta_{\rm p}
  \sin(\varphi-\varphi_{\rm p})
\\
  &+
  b_{J({\rm dc})}
  \left[
    \sin\theta_{\rm p}
    \cos\theta
    \cos(\varphi-\varphi_{\rm p})
    -
    \cos\theta_{\rm p}
    \sin\theta
  \right]
  =
  0,
  \label{eq:eq_condition_1}
\end{split}
\end{equation}
\begin{equation}
\begin{split}
  &
  H_{\rm appl}
  \sin\theta_{H}
  \sin(\varphi-\varphi_{H})
  -
  4\pi M
  \left(
    \tilde{N}_{x}
    -
    \tilde{N}_{y}
  \right)
  \sin\theta
  \sin\varphi
  \cos\varphi
\\
  &-
  a_{J({\rm dc})}
  \left[
    \sin\theta_{\rm p}
    \cos\theta
    \cos(\varphi-\varphi_{\rm p})
    -
    \cos\theta_{\rm p}
    \sin\theta
  \right]
\\
  &-
  b_{J({\rm dc})}
  \sin\theta_{\rm p}
  \sin(\varphi-\varphi_{\rm p})
  =
  0.
  \label{eq:eq_condition_2}
\end{split}
\end{equation}
In the absence of the direct current, 
$(\theta,\varphi)$ satisfying Eqs. (\ref{eq:eq_condition_1}) and (\ref{eq:eq_condition_2}) correspond to 
the equilibrium state, 
i.e., the minimum state of the energy density $E$. 
The transformation from the $xyz$-coordinate to the $XYZ$-coordinate is performed by 
multiplying the following rotation matrix to Eq. (\ref{eq:LLG}): 
\begin{equation}
  \mathsf{R}
  =
  \begin{pmatrix}
    \cos\theta & 0 & -\sin\theta \\
    0 & 1 & 0 \\
    \sin\theta & 0 & \cos\theta 
  \end{pmatrix}
  \begin{pmatrix}
    \cos\varphi & \sin\varphi & 0 \\
    -\sin\varphi & \cos\varphi & 0 \\
    0 & 0 & 1
  \end{pmatrix}.
  \label{eq:rotation_matrix}
\end{equation}
For example, 
the components of $\mathbf{p}$ in the $XYZ$-coordinate can be expressed as 
\begin{equation}
  \begin{pmatrix}
    p_{X} \\
    p_{Y} \\
    p_{Z}
  \end{pmatrix}
  =
  \begin{pmatrix}
    \cos\theta\sin\theta_{\rm p}\cos(\varphi-\varphi_{\rm p}) - \sin\theta\cos\theta_{\rm p} \\
    -\sin\theta_{\rm p}\sin(\varphi-\varphi_{\rm p}) \\
    \sin\theta\sin\theta_{\rm p}\cos(\varphi-\varphi_{\rm p}) + \cos\theta\cos\theta_{\rm p}
  \end{pmatrix}.
  \label{eq:vector_p}
\end{equation}


The alternating current exerts a small amplitude oscillation of the magnetization 
around the $Z$-axis. 
Then, the LLG equation can be linearized 
by assuming $m_{Z} \simeq 1$ and $|m_{X}|,|m_{Y}| \ll 1$, 
and is given by 
\begin{equation}
\begin{split}
  &
  \frac{1}{\gamma}
  \frac{{\rm d}}{{\rm d}t}
  \begin{pmatrix}
    m_{X} \\
    m_{Y} 
  \end{pmatrix}
  +
  \begin{pmatrix}
    -\mathcal{H}_{YX} + \alpha \mathcal{H}_{X} & \mathcal{H}_{Y} - \alpha \mathcal{H}_{XY} \\
    -\mathcal{H}_{X} - \alpha \mathcal{H}_{YX} & \mathcal{H}_{XY} + \alpha \mathcal{H}_{Y} 
  \end{pmatrix}
  \begin{pmatrix}
    m_{X} \\
    m_{Y}
  \end{pmatrix}
\\
  &=
  \begin{pmatrix}
    -a_{J({\rm ac})} p_{X} + b_{J({\rm ac})} p_{Y} \\
    -a_{J({\rm ac})} p_{Y} - b_{J({\rm ac})} p_{X} 
  \end{pmatrix},
  \label{eq:LLG_linear}
\end{split}
\end{equation}
where we use the approximation that $1+\alpha^{2}\simeq 1$ \cite{oogane06}. 
The components of $\mathcal{H}$ are defined as 
\begin{equation}
  \mathcal{H}_{X}
  =
  H_{X}
  +
  b_{J({\rm dc})}
  p_{Z},
  \label{eq:H_X_renormalize}
\end{equation}
\begin{equation}
  \mathcal{H}_{Y}
  =
  H_{Y}
  +
  b_{J({\rm dc})}
  p_{Z},
  \label{eq:H_Y_renormalize}
\end{equation}
\begin{equation}
  \mathcal{H}_{XY}
  =
  H_{XY}
  -
  a_{J({\rm dc})}
  p_{Z},
  \label{eq:H_XY_renormalize}
\end{equation}
\begin{equation}
  \mathcal{H}_{YX}
  =
  H_{YX}
  +
  a_{J({\rm dc})}
  p_{Z}.
  \label{eq:H_YX_renormalize}
\end{equation}
Here, $H_{X}=H_{ZZ}-H_{XX}$, $H_{Y}=H_{ZZ}-H_{YY}$. 
The field $H_{ij}$ ($i,j=X,Y,Z$) are the $i$-components of 
the magnetic field in the $XYZ$-coordinate proportional to $m_{j}$, 
$\mathbf{H}=(H_{XX}m_{X}+H_{XY}m_{Y},H_{YX}m_{X}+H_{YY}m_{Y},H_{ZZ}+H_{ZX}m_{X}+H_{ZY}m_{Y})$, 
where the explicit forms of $H_{ij}$ are given by 
\begin{equation}
  H_{XX}
  =
  -4\pi M 
  \left[
    \left(
      \tilde{N}_{x}
      \cos^{2}\varphi
      +
      \tilde{N}_{y}
      \sin^{2}\varphi
    \right)
    \cos^{2}\theta
    +
    \tilde{N}_{z}
    \sin^{2}\theta
  \right],
  \label{eq:H_XX}
\end{equation}
\begin{equation}
  H_{XY}
  =
  H_{YX}
  =
  4\pi M
  \left(
    \tilde{N}_{x}
    -
    \tilde{N}_{y}
  \right)
  \cos\theta
  \sin\varphi
  \cos\varphi,
  \label{eq:H_XY}
\end{equation}
\begin{equation}
  H_{YY}
  =
  -4\pi M
  \left(
    \tilde{N}_{x}
    \sin^{2}\varphi
    +
    \tilde{N}_{y}
    \cos^{2}\varphi
  \right),
  \label{eq:H_YY}
\end{equation}
\begin{equation}
  H_{ZX}
  =
  -4\pi M
  \left(
    \tilde{N}_{x}
    \cos^{2}\varphi
    +
    \tilde{N}_{y}
    \sin^{2}\varphi
    -
    \tilde{N}_{z}
  \right)
  \sin\theta
  \cos\theta,
  \label{eq:H_ZX}
\end{equation}
\begin{equation}
  H_{ZY}
  =
  -4 \pi M 
  \left(
    \tilde{N}_{y}
    -
    \tilde{N}_{x}
  \right)
  \sin\theta
  \sin\varphi
  \cos\varphi,
  \label{eq:H_ZY}
\end{equation}
\begin{equation}
\begin{split}
  H_{ZZ}
  =&
  H_{\rm appl}
  \left[
    \sin\theta_{H}
    \sin\theta
    \cos(\varphi-\varphi_{H})
    +
    \cos\theta_{H}
    \cos\theta
  \right]
\\
  &-
  4\pi M
  \left[
    \left(
      \tilde{N}_{x}
      \cos^{2}\varphi
      +
      \tilde{N}_{y}
      \sin^{2}\varphi
    \right)
    \sin^{2}\theta
    +
    \tilde{N}_{z}
    \cos^{2}\theta
  \right].
  \label{eq:H_ZZ}
\end{split}
\end{equation}


By assuming that the alternating current is given by $I_{\rm ac}\sin (2\pi ft)$, 
the solutions to $(m_{X},m_{Y})$ in Eq. (\ref{eq:LLG_linear}) are, respectively, given by \cite{comment1} 
\begin{equation}
\begin{split}
  m_{X}
  \simeq&
  {\rm Im}
  \left[
    \frac{\tilde{\gamma}(if + \tilde{\gamma}\mathcal{H}_{XY})p_{X} - \tilde{\gamma}^{2} \mathcal{H}_{Y}p_{Y}}{f^{2}-f_{\rm res}^{2}-if \Delta f}
    e^{2\pi ift}
  \right]
  \tilde{a}_{J({\rm ac})}
\\
  &-
  {\rm Im}
  \left[
    \frac{\tilde{\gamma}(if + \tilde{\gamma}\mathcal{H}_{XY})p_{Y} + \tilde{\gamma}^{2} \mathcal{H}_{Y}p_{X}}{f^{2}-f_{\rm res}^{2}-if \Delta f}
    e^{2\pi ift}
  \right]
  \tilde{b}_{J({\rm ac})},
  \label{eq:mX}
\end{split}
\end{equation}
\begin{equation}
\begin{split}
  m_{Y}
  \simeq&
  {\rm Im}
  \left[
    \frac{\tilde{\gamma}(if - \tilde{\gamma}\mathcal{H}_{YX})p_{Y} - \tilde{\gamma}^{2} \mathcal{H}_{X}p_{X}}{f^{2}-f_{\rm res}^{2}-if \Delta f}
    e^{2\pi ift}
  \right]
  \tilde{a}_{J({\rm ac})}
\\
  &-
  {\rm Im}
  \left[
    \frac{-\tilde{\gamma}(if - \tilde{\gamma}\mathcal{H}_{YX})p_{X} + \tilde{\gamma}^{2} \mathcal{H}_{X}p_{Y}}{f^{2}-f_{\rm res}^{2}-if \Delta f}
    e^{2\pi ift}
  \right]
  \tilde{b}_{J({\rm ac})},
  \label{eq:mY}
\end{split}
\end{equation}
where $\tilde{\gamma}=\gamma/(2\pi)$, 
and $\tilde{a}_{J({\rm ac})}$ and $\tilde{b}_{J({\rm ac})}$ are defined as 
$a_{J({\rm ac})}=\tilde{a}_{J({\rm ac})}\sin (2\pi ft)$ and $b_{J({\rm ac})}=\tilde{b}_{J({\rm ac})}\sin (2\pi ft)$, 
respectively. 
The resonance frequency $f_{\rm res}$ and the linewidth $\Delta f$ are, respectively, given by 
\begin{equation}
  f_{\rm res}
  =
  \frac{\gamma}{2\pi}
  \sqrt{
    \mathcal{H}_{X}
    \mathcal{H}_{Y}
    -
    \mathcal{H}_{XY}
    \mathcal{H}_{YX}
  },
  \label{eq:res}
\end{equation}
\begin{equation}
  \Delta f
  =
  \frac{\gamma}{2\pi}
  \left[
    \alpha
    \left(
      \mathcal{H}_{X}
      +
      \mathcal{H}_{Y}
    \right)
    +
    \mathcal{H}_{XY}
    -
    \mathcal{H}_{YX}
  \right].
  \label{eq:linewidth}
\end{equation}
In the absence of the direct current, 
Eq. (\ref{eq:res}) is the ferromagnetic resonance (FMR) frequency, 
$f_{\rm FMR}=\gamma \sqrt{H_{X}H_{Y}-H_{XY}^{2}}/(2\pi)$. 
Since $\beta$ is on the order of 0.1 \cite{tulapurkar05,kubota08,sankey08}, 
and $a_{J({\rm dc})}$ is on the order of a small parameter $\alpha$, 
$\alpha b_{J({\rm dc})}$ in Eq. (\ref{eq:linewidth}) is negligible. 
Thus, $\Delta f$ can be approximated to 
\begin{equation}
  \Delta f
  \simeq 
  \frac{\gamma}{2\pi}
  \left[
    \alpha
    \left(
      H_{X}
      +
      H_{Y}
    \right)
    - 
    2 a_{J({\rm dc})}
    p_{Z}
  \right].
\end{equation}



\section{Spin torque diode voltage}
\label{sec:Spin torque diode voltage}

The magnetoresistance of an MTJ is given by 
$R=R_{\rm P}+(\Delta R/2)(1-\mathbf{m}\cdot\mathbf{p})$, 
where $\Delta R=R_{\rm AP}-R_{\rm P}$ is the difference in the resistances 
between the parallel ($R_{\rm P}$) and the anti-parallel ($R_{\rm AP}$) alignments 
of the magnetizations. 
The spin torque diode voltage is given by  
$V_{\rm dc}=T^{-1} \int_{0}^{T} I(t) R(t) {\rm d}t$, 
where $T=1/f$ is the period of the alternating current. 
By using Eqs. (\ref{eq:mX}) and (\ref{eq:mY}), 
the explicit form of the spin torque diode voltage is given by 
\begin{equation}
\begin{split}
  V_{\rm dc}
  &=
  \frac{\Delta R I_{\rm ac}}{4}
  {\rm Re}
  \left[
    \frac{-[if(p_{X}^{2}+p_{Y}^{2})+\tilde{\gamma}\mathscr{H}_{a}]\tilde{\gamma} \tilde{a}_{J({\rm ac})} + \tilde{\gamma}^{2} \mathscr{H}_{b} \tilde{b}_{J({\rm ac})}}
      {f^{2}-f_{\rm res}^{2}-if \Delta f}
  \right]
\\
  &=
  \frac{\Delta R I_{\rm ac}}{4}
  \left[
    \mathscr{L}(f)
    +
    \mathscr{A}(f)
  \right], 
  \label{eq:voltage}
\end{split}
\end{equation}
where $\mathscr{H}_{a}$ and $\mathscr{H}_{b}$ are, respectively, given by 
\begin{equation}
\begin{split}
  \mathscr{H}_{a}
  =&
  \mathcal{H}_{XY}
  p_{X}^{2}
  -
  \mathcal{H}_{YX}
  p_{Y}^{2}
  +
  \left(
    \mathcal{H}_{X}
    -
    \mathcal{H}_{Y}
  \right)
  p_{X}
  p_{Y},
  \label{eq:H_a}
\end{split}
\end{equation}
\begin{equation}
\begin{split}
  \mathscr{H}_{b}
  =
  \mathcal{H}_{Y}
  p_{X}^{2}
  +
  \mathcal{H}_{X}
  p_{Y}^{2}
  +
  \left(
    \mathcal{H}_{XY}
    +
    \mathcal{H}_{YX}
  \right)
  p_{X}
  p_{Y}.
  \label{eq:H_b}
\end{split}
\end{equation}
The Lorentzian and the anti-Lorentzian parts, $\mathscr{L}(f)$ and $\mathscr{A}(f)$ are, 
respectively given by 
\begin{equation}
  \mathscr{L}(f)
  =
  \frac{f^{2} \Delta f \tilde{\gamma} \tilde{a}_{J({\rm ac})} (1-p_{Z}^{2})}{(f^{2}-f_{\rm res}^{2})^{2} + (f \Delta f)^{2}},
  \label{eq:Lorentzian}
\end{equation}
\begin{equation}
  \mathscr{A}(f)
  =
  -\frac{(f^{2}-f_{\rm res}^{2}) \tilde{\gamma}^{2} (\mathscr{H}_{a} \tilde{a}_{J({\rm ac})} - \mathscr{H}_{b} \tilde{b}_{J({\rm ac})})}{(f^{2}-f_{\rm res}^{2})^{2} + (f \Delta f)^{2}}.
  \label{eq:anti_Lorentzian}
\end{equation}
As shown, 
the Lorentzian part depends on the Slonczewski torque only 
while the anti-Lorentzian part depends on both the Slonczewski torque and the field like torque, 
in general. 


The peak of the spin torque diode voltage appears 
around the resonance frequency, $f_{\rm res}$, 
where the Lorentzian part shows a peak 
while the anti-Lorentzian part is zero. 
At $f=f_{\rm res}$, the spin torque diode voltage is 
\begin{equation}
  V_{\rm dc}(f_{\rm res})
  =
  \frac{\Delta R I_{\rm ac}}{4}
  \frac{\tilde{a}_{J({\rm ac})}\sin^{2}\psi}{\alpha (H_{X} + H_{Y}) - 2 a_{J({\rm dc})} \cos\psi}, 
  \label{eq:voltage_tmp}
\end{equation}
where $\psi=\cos^{-1}p_{Z}=\cos^{-1}\mathbf{m}^{(0)}\cdot\mathbf{p}$ 
in the relative angle between the magnetizations of the free and the pinned layers. 
Equation (\ref{eq:voltage_tmp}) is maximized 
when the relative angle of the magnetizations is given by 
\begin{equation}
  \psi^{\rm opt}
  =
  \cos^{-1}
  \left[
    \frac{I_{\rm c}}{I_{\rm dc}}
    \mp
    \sqrt{
      \left(
        \frac{I_{\rm c}}{I_{\rm dc}}
      \right)^{2}
      -
      1
    }
  \right],
  \label{eq:p_z_opt}
\end{equation}
where the double sign "$\mp$" means 
the upper ($-$) for $I_{\rm dc}/I_{\rm c}>0$ 
and the lower ($+$) for $I_{\rm dc}/I_{\rm c}<0$. 
The critical current of the spin torque induced magnetization dynamics 
in the case of $\mathbf{m}^{(0)} \parallel \mathbf{p}$, $I_{\rm c}$, is given by  
\begin{equation}
  I_{\rm c}
  =
  \frac{2 \alpha eMV}{\hbar \eta}
  \left(
    \frac{H_{X} + H_{Y}}{2}
  \right).
  \label{eq:Ic}
\end{equation}
Since $\psi$ is a real number, 
the following condition should be satisfied: 
\begin{equation}
  \bigg|
    \frac{I_{\rm c}}{I_{\rm dc}}
  \bigg|
  >
  1.
  \label{eq:opt_condition_2}
\end{equation}
This condition means that 
the linear approximation cannot be applied to the LLG equation 
when the spin torque overcomes the damping. 
The maximized spin torque diode voltage is given by 
\begin{equation}
  V_{\rm dc}^{\rm opt}(f_{\rm res})
  =
  \frac{\Delta R I_{\rm ac}^{2}}{4 I_{\rm dc}}
  \left[
    \frac{I_{\rm c}}{I_{\rm dc}}
    \mp
    \sqrt{
      \left(
        \frac{I_{\rm c}}{I_{\rm dc}}
      \right)^{2}
      -
      1
    }
  \right]. 
  \label{eq:voltage_max}
\end{equation}
Equations (\ref{eq:p_z_opt}) and (\ref{eq:voltage_max}) are the main results 
in this section, 
and can be regarded as generalizations 
of the result in Ref. \cite{taniguchi13}. 
We emphasize that 
the optimum condition, Eq. (\ref{eq:p_z_opt}), depends 
on not only the material (sample) parameters and the applied field 
but also the magnitude and direction of the direct current. 
It should be noted that Eq. (\ref{eq:p_z_opt}) is $90^{\circ}$ for $I_{\rm dc}=0$, 
and shifts from this orthogonal alignment for a finite $I_{\rm dc}$. 


Equation (\ref{eq:p_z_opt}) is the optimum condition of the magnetization alignment 
to maximize the spin torque diode voltage. 
However, in experiments, 
the direction of the applied field is more easily controlled, than 
the magnetization alignment, 
because the direction of the magnetization of the pinned layer is fixed by 
the exchange bias from an anti-ferromagnetic layer. 
In the next section, by using Eq. (\ref{eq:p_z_opt}), 
we derive the analytical formula of the optimum condition of the applied field direction 
to maximize the spin torque diode voltage 
in an MTJ with a perpendicularly magnetized free layer 
and an in-plane magnetized pinned layer. 



\section{Optimum condition of applied field direction}
\label{sec:Optimum condition of applied field direction}

In an MTJ with a perpendicularly magnetized free layer and 
an in-plane magnetized pinned layer, 
$(\theta_{\rm p},\varphi_{\rm p})$ in Fig. \ref{fig:fig1} 
are $(90^{\circ},0^{\circ}$). 
The free layer has uniaxial anisotropy along the easy axis 
which is normal to the film plane, 
and has a circular cross section. 
The components of the anisotropy field are 
$4\pi M \tilde{N}_{x}=4\pi M \tilde{N}_{y}=0$, 
and $4\pi M \tilde{N}_{z}=-H_{\rm K}+4\pi M$, 
where the $z$-axis is parallel to the easy axis. 
Since we are interested in the perpendicularly magnetized free layer, 
the anisotropy field $H_{\rm K}$ should be larger than the demagnetization field $4\pi M$. 
The $x$-axis is parallel to the magnetization of the pinned layer. 
We assume that the magnetic field is applied 
tilted from the $z$-axis with the angle $\theta_{H}(0 < \theta_{H} < \pi)$. 
In the following, 
we investigate the optimum direction of the applied field in the film-plane, $\varphi_{H}$, 
to maximize the diode voltage. 





We assume that 
the steady state $(\theta,\varphi)$ is determined by Eqs. (\ref{eq:eq_condition_1}) and (\ref{eq:eq_condition_2}) 
by neglecting the spin torque term, 
i.e., $(\theta,\varphi)$ corresponds to the equilibrium state, 
because the spin torque term is on the order of a small parameter $\alpha$. 
The equilibrium state of the free layer satisfies 
\begin{equation}
  H_{\rm appl}
  \sin(\theta-\theta_{H})
  +
  \left(
    H_{\rm K}
    -
    4\pi M
  \right)
  \sin\theta
  \cos\theta
  =
  0,
  \label{eq:eq_condition_perp}
\end{equation}
and $\varphi=\varphi_{H}$. 
Then, the critical current, 
\begin{equation}
  I_{\rm c}
  =
  \frac{2 \alpha eMV}{\hbar \eta}
  \left[
    H_{\rm appl}
    \cos(\theta-\theta_{H})
    +
    H_{\perp}
    \frac{\cos^{2}\theta + \cos 2 \theta}{2}
  \right],
  \label{eq:Ic_perp}
\end{equation}
is independent of $\varphi_{H}$. 
Equation (\ref{eq:p_z_opt}) can be expressed as 
\begin{equation}
  \sin\theta
  \cos\varphi
  =
  \frac{I_{\rm c}}{I_{\rm dc}}
  \mp
  \sqrt{
    \left(
      \frac{I_{\rm c}}{I_{\rm dc}}
    \right)^{2}
    -
    1
  }.
  \label{eq:p_z_opt_perp_free_in_plane_pin}
\end{equation}
As mentioned above, 
$\varphi$ on the left-hand side of Eq. (\ref{eq:p_z_opt_perp_free_in_plane_pin}) can be replaced by $\varphi_{H}$. 
When the condition 
\begin{equation}
  \Bigg|
    \frac{1}{\sin\theta}
    \left[
      \frac{I_{\rm c}}{I_{\rm dc}}
      \mp
      \sqrt{
        \left(
          \frac{I_{\rm c}}{I_{\rm dc}}
        \right)^{2}
        -
        1
      }
    \right]
  \Bigg|
  <1
  \label{eq:condition_perp_1}
\end{equation}
is satisfied, 
the diode voltage is maximized at the field direction 
\begin{equation}
  \varphi_{H}
  =
  \cos^{-1}
  \left\{
    \frac{1}{\sin\theta}
    \left[
      \frac{I_{\rm c}}{I_{\rm dc}}
      \mp
      \sqrt{
        \left(
          \frac{I_{\rm c}}{I_{\rm dc}}
        \right)^{2}
        -
        1
      }
    \right]
  \right\}.
  \label{eq:p_z_opt_perp}
\end{equation}
Equation (\ref{eq:p_z_opt_perp}) is the main result in this paper. 
For a relatively small applied field magnitude, 
the equilibrium state is close to the easy axis, i.e., $\theta \simeq \sin\theta \ll 1$, 
and Eq. (\ref{eq:condition_perp_1}) is not satisfied. 
Then, the diode voltage is maximized at $\varphi_{H}=0$ or $\pi$, 
depending on the direction of the current. 
However, by increasing the applied field magnitude, 
the magnetization tilts from the easy axis, 
and Eq. (\ref{eq:condition_perp_1}) is satisfied. 
The optimum direction of the applied magnetic field then 
shifts from $\varphi_{H}=0,\pi$ to $\varphi_{H}$ given by Eq. (\ref{eq:p_z_opt_perp}). 


\begin{figure}
\centerline{\includegraphics[width=1.0\columnwidth]{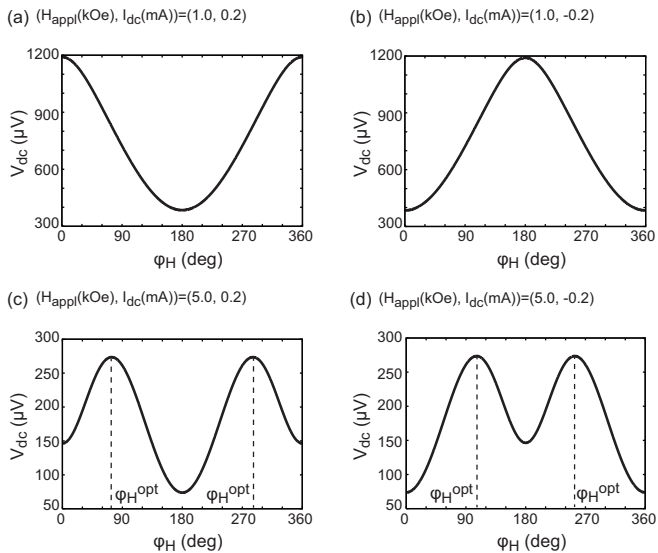}}\vspace{-3.0ex}
\caption{
         The dependence of the diode voltage at the resonance, $V_{\rm dc}(f_{\rm res})$, 
         on the applied field direction, $\varphi_{H}$. 
         The values of the applied field and the current are 
         (a) $(H_{\rm appl}({\rm kOe}),I_{\rm dc}({\rm mA}))=(1.0,0.2)$, 
         (b) $(1.0,-0.2)$, 
         (c) $(5.0,0.2)$, 
         and (d) $(5.0,-0.2)$, respectively. 
         \vspace{-3ex}}
\label{fig:fig3}
\end{figure}


The physical meaning of the condition (\ref{eq:condition_perp_1}) is as follows. 
As mentioned after Eq. (\ref{eq:voltage_max}), 
the spin torque diode voltage is maximized near the orthogonal alignment of the magnetization. 
When the magnitude of the applied field is small, 
this condition is approximately satisfied. 
Then, the magnetization should oscillate in the $xz$-plane 
to obtain a large oscillation amplitude of TMR 
because $\mathbf{p}$ points to the $x$-direction. 
Thus, Eq. (\ref{eq:p_z_opt_perp}) is $0$ or $\pi$. 
However, 
the magnetization moves to the $xy$-plane for a relatively large applied field. 
To keep the relative angle of the magnetizations close to Eq. (\ref{eq:p_z_opt}), 
the magnetization of the free layer should shift from the $x$-axis 
by changing the field direction. 
Thus, the optimum field direction shifts from $\varphi_{H}=0,\pi$, 
according to Eq. (\ref{eq:p_z_opt_perp}). 


The reason why the analytical solution of the optimum applied field direction, 
Eq. (\ref{eq:p_z_opt_perp}), can be obtained in this system 
is that, because of the axial symmetry, 
$\varphi$ in Eq. (\ref{eq:p_z_opt_perp_free_in_plane_pin}) can be 
replaced by $\varphi_{H}$. 
In the general system, 
both sides of Eq. (\ref{eq:p_z_opt}) depend on the applied field direction $(\theta_{H},\varphi_{H})$ 
through Eqs. (\ref{eq:eq_condition_1}) and (\ref{eq:eq_condition_2}). 
Consequently, an analytical expression of 
the optimum field direction cannot be obtained. 


Let us quantitatively estimate the optimum direction, $\varphi_{H}$. 
Figure \ref{fig:fig3} shows 
the dependence of the spin torque diode voltage, $V_{\rm dc}(f_{\rm res})$, 
on the applied field direction, $\varphi_{H}$, for several values of $H_{\rm appl}$ and $I_{\rm dc}$. 
The values of the parameters \cite{kubota13} are 
$M=1313$ emu/c.c., 
$H_{\rm K}=17.9$ kOe, 
$\theta_{H}=60^{\circ}$, 
$V=\pi \times 50 \times 50 \times 2$ nm${}^{3}$, 
$\gamma=17.32$ MHz/Oe, 
$\alpha=0.005$, 
$\eta=0.33$, 
$\beta=0.1$, 
$I_{\rm ac}=0.1$ mA, and 
$\Delta R=100$ $\Omega$, respectively. 
The values of $H_{\rm appl}$ and $I_{\rm dc}$ are 
(a) $(H_{\rm appl}({\rm kOe}),I_{\rm dc}({\rm mA}))=(1.0,0.2)$, 
(b) $(1.0,-0.2)$, 
(c) $(5.0,0.2)$, 
and (d) $(5.0,-0.2)$, respectively,
where the value of $I_{\rm dc}$ is chosen 
to observe the shift of the optimum $\varphi_{H}$ from $0$ or $\pi$ to a ceratin angle 
in a typical range of $H_{\rm appl}$ in experiments \cite{kubota13}. 
The current magnitude (0.2 mA) is also a typical value used in experiments (for example, Ref. \cite{kubota08}). 
The steady state of the magnetization of the free layer is 
$\theta=26.3^{\circ}$ for $H_{\rm appl}=1.0$ kOe. 
In this case, Eq. (\ref{eq:condition_perp_1}) is not satisfied, 
and thus, the diode voltage is maximized at $\varphi_{H}=0$ for $I_{\rm dc}/I_{\rm c}>0$ 
and at $\varphi_{H}=\pi$ for $I_{\rm dc}/I_{\rm c}<0$, 
as shown in Figs. \ref{fig:fig3} (a) and (b), respectively. 
On the other hand, 
the steady state is given by $\theta=52.0^{\circ}$ for $H_{\rm appl}=5.0$ kOe. 
The condition, Eq. (\ref{eq:condition_perp_1}), is satisfied, 
and the optimum direction of the applied field is given by 
$\varphi_{H}=63.7^{\circ}$ and $296.3^{\circ}$ for $I_{\rm dc}=0.2$ mA 
and $73.9^{\circ}$ and $106.1^{\circ}$ for $-0.2$ mA, 
respectively. 
The maximized voltage is estimated to be 272 $\mu$V 
while the diode voltages at $\varphi_{H}=0$ and $\pi$ for $I_{\rm dc}>0$ are 
estimated to be 146 and 73 $\mu$V, respectively. 
Since the relative angle between the magnetizations 
decreases as the applied field magnitude increases, 
the maximized voltage for $H_{\rm appl}=5.0$ kOe is smaller than 
that for $H_{\rm appl}=1.0$ kOe. 


\begin{figure}
\centerline{\includegraphics[width=0.7\columnwidth]{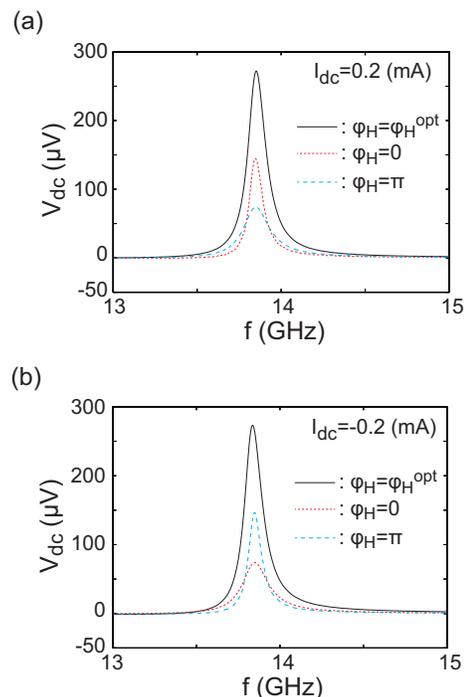}}\vspace{-3.0ex}
\caption{
         The dependences of the spin torque diode voltage at $\varphi_{H}=\varphi_{H}^{\rm opt}$ (black, solid),$0$ (red, dotted), and $\pi$ (blue, dashed) 
         on the frequency of the applied alternating current. 
         The values of the direct current are (a) $I_{\rm dc}=0.2$ mA and (b) $-0.2$ mA, respectively. 
         \vspace{-3ex}}
\label{fig:fig4}
\end{figure}



We perform numerical simulations \cite{taniguchi11} 
to confirm the above analytical results. 
Figures \ref{fig:fig4} (a) and (b) shows 
the dependences of the spin torque diode voltage at $\varphi_{H}=\varphi_{H}^{\rm opt},0$, and $\pi$ 
on the frequency of the alternating current 
with $I_{\rm dc}=0.2$ mA and $-0.2$ mA, respectively. 
The magnitude of the applied magnetic field is $H_{\rm appl}=5.0$ kOe. 
A sharp peak of the diode voltage appears near the FMR frequency, $f_{\rm FMR}\simeq 13.8$ GHz. 
The magnitudes of the diode voltage  at $f=f_{\rm res}$ agree well 
with the results shown in Fig. \ref{fig:fig3}, 
demonstrating the validity of the above analytical formula. 





\section{Conclusions}
\label{sec:Conclusions}

In conclusion, 
we derive the optimum condition of the applied field direction 
to maximize the diode voltage of 
an MTJ with a perpendicularly magnetized free layer 
and an in-plane magnetized pinned layer, 
which was recently developed in experiments. 
For a relatively small applied field, 
the diode voltage is maximized 
when the projection of the applied field to the film-plane is 
parallel or anti-parallel to the magnetization of the pinned layer. 
However, the voltage is maximized at a certain direction 
shifted from the parallel or anti-parallel direction 
by increasing the applied field magnitude. 
These results are confirmed by numerically solving the Landau-Lifshitz-Gilbert equation. 



The authors would like to acknowledge 
H. Kubota, H. Maehara, A. Emura, T. Yorozu, H. Arai, S. Yuasa,  K. Ando, and S. Miwa 
for the valuable discussions they had with us. 
This work was supported by JSPS KAKENHI Number 23226001. 


\appendix


\section{Spin torque diode voltage and its maximized condition with finite $\lambda$}

In this appendix, 
the spin torque diode voltage 
with finite $\lambda$ in Eq. (\ref{eq:factor_g}) is derived. 


First, let us briefly describe the importance of $\lambda$, 
which arises from the dependence of the tunneling probability 
on the magnetization alignment \cite{suzuki09}. 
Since the magnitude of $\lambda$ is small, 
for simplicity, 
we assume $\lambda$ is zero in some cases \cite{taniguchi11}. 
However, when the magnetization alignment of the free and the pinned layers in equilibrium is orthogonal 
($\mathbf{m}^{(0)} \perp \mathbf{p}$), 
a finite $\lambda$ plays a key role in the spin torque induced 
magnetization dynamics. 
For example, 
when $\lambda$ is neglected, 
the critical current for the magnetization dynamics, 
Eq. (\ref{eq:Ic_general}) shown below, 
diverges for $\mathbf{m}^{(0)}\perp \mathbf{p}$ (i.e., $p_{Z}=0$). 
This is because the work done by spin torque is zero 
at this alignment, 
and thus, the spin torque cannot overcome the damping. 
However, if $\lambda \neq 0$, 
the critical current remains finite 
because the work done by spin torque is also finite, 
and can be larger than the energy dissipation due to the damping \cite{taniguchi13a}. 


Now we calculate the diode voltage with finite $\lambda$. 
Let us redefine $a_{J}$ and $b_{J}$ as 
\begin{equation}
  a_{J}
  \equiv
  \frac{\hbar \eta I}{2eMV (1 + \lambda p_{Z})},
  \label{eq:a_J_lambda}
\end{equation}
and $b_{J}=\beta a_{J}$. 
Also, we introduce $\varLambda$ as 
\begin{equation}
  \varLambda
  \equiv
  \frac{\lambda}{1 + \lambda p_{Z}}.
\end{equation}
Then, instead of Eqs. (\ref{eq:H_X_renormalize})-(\ref{eq:H_YX_renormalize}), 
we redefine $\mathcal{H}_{X}$, $\mathcal{H}_{Y}$, $\mathcal{H}_{XY}$, and $\mathcal{H}_{YX}$ as 
\begin{equation}
  \mathcal{H}_{X}
  =
  H_{X}
  +
  b_{J({\rm dc})}
  p_{Z}
  +
  a_{J({\rm dc})} 
  \varLambda
  p_{X}
  p_{Y}
  +
  b_{J({\rm dc})}
  \varLambda
  p_{X}^{2},
  \label{eq:H_X_redefine}
\end{equation}
\begin{equation}
  \mathcal{H}_{Y}
  =
  H_{Y}
  +
  b_{J({\rm dc})}
  p_{Z}
  -
  a_{J({\rm dc})}
  \varLambda
  p_{X}
  p_{Y}
  +
  b_{J({\rm dc})}
  \varLambda
  p_{Y}^{2},
  \label{eq:H_Y_redefine}
\end{equation}
\begin{equation}
  \mathcal{H}_{XY}
  =
  H_{XY}
  -
  a_{J({\rm dc})}
  p_{Z}
  -
  a_{J({\rm dc})}
  \varLambda
  p_{Y}^{2}
  -
  b_{J({\rm dc})}
  \varLambda
  p_{X}
  p_{Y},
  \label{eq:H_XY_redefine}
\end{equation}
\begin{equation}
  \mathcal{H}_{YX}
  =
  H_{YX}
  +
  a_{J({\rm dc})}
  p_{Z}
  +
  a_{J({\rm dc})}
  \varLambda
  p_{X}^{2}
  -
  b_{J({\rm dc})}
  \varLambda
  p_{X}
  p_{Y}.
  \label{eq:H_YX_redefine}
\end{equation}
By using these $\mathcal{H}$, 
the resonance frequency, the linewidth, and $\mathscr{H}$ 
are redefined according to Eqs. (\ref{eq:res}), (\ref{eq:linewidth}), (\ref{eq:H_a}), and (\ref{eq:H_b}), respectively. 
Then, the diode voltage is given by 
Eqs. (\ref{eq:voltage}), (\ref{eq:Lorentzian}), and (\ref{eq:anti_Lorentzian}). 
The critical current for the magnetization dynamics is given by 
\begin{equation}
  \mathscr{I}_{\rm c}
  =
  \frac{2\alpha eMV}{\hbar \eta (1-\alpha \beta) [p_{Z} + \varLambda (1-p_{Z}^{2})]}
  \left(
    \frac{H_{X} + H_{Y}}{2}
  \right).
  \label{eq:Ic_general}
\end{equation}


It is difficult for an arbitrary $\lambda$ ($-1 < \lambda < 1$) to derive the optimum condition. 
However, the optimum condition for $|\lambda|\ll 1$ can be derived as follows. 
In this case, the diode voltage at $f=f_{\rm res}$ is given by 
Eq. (\ref{eq:voltage_tmp}) 
in which $a_{J}$ is replaced by Eq. (\ref{eq:a_J_lambda}). 
Then, the diode voltage is maximized 
when the relative angle of $\mathbf{m}^{(0)}$ and $\mathbf{p}$ is given by 
\begin{equation}
  \psi^{\rm opt} 
  =
  \cos^{-1}
  \left\{
    \frac{(I_{\rm c}/I_{\rm dc}) \mp \sqrt{(I_{\rm c}/I_{\rm dc})^{2} - [1- \lambda(I_{\rm c}/I_{\rm dc})]^{2}}}
      {1 - \lambda (I_{\rm c}/I_{\rm dc})}
  \right\},
  \label{eq:p_z_opt_lambda}
\end{equation}
where $I_{\rm c}$ is defined by Eq. (\ref{eq:Ic}). 
Equation (\ref{eq:p_z_opt_lambda}) is identical to Eq. (\ref{eq:p_z_opt}) 
in the limit of $\lambda \to 0$. 
The maximized voltage at $f=f_{\rm res}$ is given by 
\begin{equation}
  V_{\rm dc}^{\rm opt}
  =
  \frac{\Delta R I_{\rm ac}^{2}}{4 I_{\rm dc}}
  \left\{
    \frac{(I_{\rm c}/I_{\rm dc}) \mp \sqrt{(I_{\rm c}/I_{\rm dc})^{2} - [1- \lambda (I_{\rm c}/I_{\rm dc})]^{2}}}
      {[1 - \lambda (I_{\rm c}/I_{\rm dc})]^{2}}
  \right\}
  \label{eq:voltage_max_lambda}
\end{equation}


The spin torque diode effect is useful for estimating the value of $\lambda$ experimentally. 
For example, let us consider the spin torque diode effect 
of MTJ with the perpendicularly magnetized free layer and the in-plane magnetized pinned layer 
discussed in Sec. \ref{sec:Optimum condition of applied field direction}.
The direct current is assumed to be zero. 
By fixing the magnitude ($H_{\rm appl}$) and the tilted angle ($\theta_{H}$) of the applied field, 
the resonance frequencies and the linewidths at 
a certain $\varphi_{H}$ and $\pi-\varphi_{H}$ are identical. 
Then, the ratio of the diode voltages at $\varphi_{H}$ and $\pi-\varphi_{H}$ is given by 
\begin{equation}
  \frac{V_{\rm dc}(f=f_{\rm res},\varphi_{H})}{V_{\rm dc}(f=f_{\rm res},\pi-\varphi_{H})}
  =
  \frac{1 - \lambda \sin\theta}{1 + \lambda \sin\theta},
\end{equation}
where the factor $1 \mp \lambda \sin\theta$ appears 
from $\tilde{a}_{J({\rm ac})} \propto 1/(1 + \lambda p_{Z})$ 
in the numerator of Eq. (\ref{eq:voltage_tmp}). 
Since the value of $\theta$ is determined by Eq. (\ref{eq:eq_condition_perp}), 
the value of $\lambda$ can be estimated by this ratio. 
This method of estimating $\lambda$ is applicable to general system 
if there are at least two equilibrium states with identical resonance frequencies and linewidths 
and different relative angles with the magnetization of the pinned layer.


\begin{thebibliography}{36}
\bibitem[{com()}]{comment1}
\bibinfo{note}{We neglect terms proportional to $\alpha a_{J({\rm ac})}$ and
  $\alpha b_{J({\rm ac)}}$ because these terms are, at least, on the order of
  $\alpha^{2}$.}
\end{thebibliography}
\end{document}